\pdfoutput=1
\RequirePackage{ifpdf}
\ifpdf 
\documentclass[pdftex]{sigma}
\else
\documentclass{sigma}
\fi

\usepackage{dsfont}
\usepackage{tikz}
\newcommand{\AND}{\qquad \mathrm{and} \qquad}

\numberwithin{equation}{section}

\newtheorem{Theorem}{Theorem}[section]
\newtheorem*{Theorem*}{Theorem}
\newtheorem{Corollary}[Theorem]{Corollary}
\newtheorem{Lemma}[Theorem]{Lemma}
\newtheorem{Proposition}[Theorem]{Proposition}

\theoremstyle{definition}
\newtheorem{Definition}[Theorem]{Definition}

\newtheorem{Example}[Theorem]{Example}
\newtheorem{Remark}[Theorem]{Remark}

\begin{document}

\allowdisplaybreaks

\newcommand{\arXivNumber}{2505.02814}

\renewcommand{\PaperNumber}{031}

\FirstPageHeading

\ShortArticleName{Characterization of Gaussian Tensor Ensembles}

\ArticleName{Characterization of Gaussian Tensor Ensembles}

\Author{R\'emi BONNIN}

\AuthorNameForHeading{R.~Bonnin}

\Address{Aix-Marseille Universit\'e, CNRS, I2M, Marseille, France}
\Email{\mail{remi.bonnin@ens.psl.eu}, \mail{remi.bonnin@univ-amu.fr}}

\ArticleDates{Received July 09, 2025, in final form March 13, 2026; Published online March 31, 2026}

\Abstract{The starting point of this work is a theorem due to Maxwell characterizing the distribution of a Gaussian vector with at least two coordinates. We define the Gaussian orthogonal, unitary and symplectic tensor ensembles for notions of real symmetric, hermitian and self-dual hermitian tensors which recover the classical vector and matrix Gaussian ensembles when the order is one and two. We give a complete family of invariant polynomials for orthogonal, unitary and symplectic transformations and we prove a Maxwell-type theorem for these Gaussian tensor distributions unifying and extending the ones known for vectors and matrices.}


\Classification{60B20; 81T32}

\section{Introduction and main results}

\subsection{Characterization of Gaussian distributions}

Gaussian distributions are ubiquitous in modern science, and there are numerous ways to observe and understand their occurrence. They are, among others, the maximum entropy at fixed variance or the eigenvector of the Fourier transform. One very nice {\em geometric} characterization of multivariate Gaussian distributions is due to Maxwell \cite{feller,maxwell}. His theorem states that a¬random vector of dimension two or more has independent entries and is {\em rotationally invariant} if and only if its components are Gaussian, centered, with the same variance. In other words, such a random vector has a law with a density of the form
\[ H \mapsto \alpha \exp\bigl(-\vert H \vert_2^2 / \gamma\bigr), \]
with $(\alpha,\gamma) \in \mathbb{R}^*_+ \times \mathbb{R}^*_+$. Indeed, the rotational invariance forces the law to depend only on the norm of the vector and the independence of the entries then forces it to be proportional to a~dilated Gaussian vector law. A short proof is given in Section \ref{vector}. It has a lot of application in physics, see the reviews \cite{Gyenis2017, robson17}, and it inspired Boltzmann for his kinetic evolution equation. Maxwell observes this phenomenon in order to derive the distribution of velocities in an ideal gas at equilibrium. When the vector has three entries, the distribution is called {\em Maxwellian} in statistical physics.

There exists a generalization of this result to the order two, due to Rosenzweig and Porter \cite{rosenzweigporter}.  The theorem states that a real symmetric (resp.\ hermitian, resp.\ self-dual hermitian) random matrix of dimension two or more has in the same time independent entries (up to symmetries) and a law invariant by orthogonal (resp.\ unitary, resp.\ symplectic) transformation if and only if its law has a density of the form
\[ H \mapsto \exp \bigl( - a \operatorname{Tr}\bigl(H^2\bigr) + b \operatorname{Tr}(H) + c \bigr),\]
where $(a,b,c) \in \mathbb{R}^*_+ \times \mathbb{R} \times \mathbb{R}$. We aim to prove that both vector and matrix results may be unified in an overall characterization of Gaussian tensor ensembles of any order $p\geq 1$.

\begin{theorem*}[main]
For any $p\geq 1$ and $N\geq 2$, a real symmetric $($resp.\ hermitian, resp.\ self-dual hermitian$)$ random tensor $H$ of order $p$ and dimension $N$ \eqref{realcomplextens} has
\begin{itemize}\itemsep=0pt
    \item[--] independent entries $($up to symmetries$)$,
    \item[--] a law invariant by orthogonal $($resp.\ unitary, resp.\ symplectic$)$ transformation $($see Definition~{\rm\ref{def:tens_inv}}$)$,
\end{itemize}
if and only if it belongs to the Gaussian orthogonal $($resp.\ unitary, resp.\ symplectic$)$ tensor ensemble \eqref{GTE}, that is, it has a law with a density of the form
\[ H \mapsto \alpha \exp \bigl( - \| H - \beta \mathcal{I} \|_{\mathrm{F}}^2 / \gamma \bigr),\]
where $(\alpha,\beta,\gamma) \in \mathbb{R}^*_+ \times \mathbb{R} \times \mathbb{R}^*_+$, and $\mathcal{I}$ is the symmetric tensor identity $($Definition {\rm\ref{def:identity}}$)$.
\end{theorem*}

This phenomenon is purely multivariate as it is always true in dimension two or more, but trivially false when $N=1$, where any law is product and rotational invariance becomes symmetry. A counterexample is given by the Rademacher law $\frac{\delta_{-1}+\delta_1}{2}$.

\subsection{Real and complex tensors}\label{realcomplextens}

A {\em tensor} is an element of the vector space $\bigl(\mathbb{C}^N\bigr)^{\otimes p}$ for some $p\geq 1$ and $N\geq 1$. The integer $p$ is called the {\em order}, when $p=1$ (resp.\ $p=2$) we recover complex vectors (resp.\ matrices) space. The integer $N$ is called the {\em dimension}, note that we are dealing in this work only with {\em cubic} tensors, that is, when each one of the $p$ legs has the same dimension $N$, because we focus on global invariance and not local one. For a complex tensor $H \in \bigl(\mathbb{C}^N\bigr)^{\otimes p}$, the {\em Frobenius norm} of~$H$ is the sum of the square of the modulus of the entries of the tensor, that is,
\[ \| H \|_{\mathrm{F}}^2= \sum_{i_1,\dots,i_p} \vert H_{i_1,\dots,i_p} \vert^2.\]
When $H$ is real, it is just the sum of the square of the entries.

{\bf Real symmetric and antisymmetric tensors.}
Let \smash{$H \in \bigl(\mathbb{R}^N\bigr)^{\otimes p}$} be a real tensor of order~$p$ and dimension $N$ for some $p\geq 1$, $N\geq 1$. The symmetric group $S_p$ acts on \smash{$\bigl(\mathbb{R}^N\bigr)^{\otimes p}$} by permutation of the indices
\smash{$ H^{\sigma}_{i_1,\dots,i_p} = H_{i_{\sigma(1)},\dots,i_{\sigma(p)}}$},
for $\sigma \in S_p$. A~tensor is said {\em symmetric} if
$H=H^{\sigma}$,
for all $\sigma \in S_p$. Note that when $p=1$, a~vector is always symmetric. We~denote~$\mathcal{S}^{(p)}(N)$ the set of real symmetric tensors of order $p$ and dimension $N$. Moreover, we define an {\em antisymmetric} tensor as a real tensor satisfying
\smash{$H^{\tau}_{i_1,\dots,i_p} = \epsilon(\tau)H_{i_1,\dots,i_p}$},
for all $\tau \in S_p$.
Similarly, we denote~$\mathcal{A}^{(p)}(N)$ the set of real antisymmetric tensors of order $p$ and dimension $N$. We say $(i_1,\dots,i_p) \sim (j_1,\dots,j_p)$ if $\forall k, i_k = j_{\sigma(k)} $ for some $\sigma \in \mathrm{S}_p$ and we denote
\[ \mathfrak{P}=\{(i_1,\dots,i_p) \in \{1,\dots,N\}^p \mid \exists j_1,\dots,j_{p/2} \text{ s.t.\ } (i_1,\dots,i_p) \sim (j_1,j_1,\dots,j_{p/2},j_{p/2}) \}, \]
the set of {\em paired} indices. When $p$ is even, one can check that if $H$ is antisymmetric, $H_{i_1,\dots,i_p}=0$ for all $(i_1,\dots,i_p) \in \mathfrak{P}$. Denote $\Gamma_{i_1,\dots,i_p}$ the cardinality of the class of $(i_1,\dots,i_p)$ under $\sim$, that~is,
\[  \Gamma_{i_1,\dots,i_p} = \frac{p!}{\prod_{j=1}^N c_j(i_1,\dots,i_p)!},\]
with $c_j(i_1,\dots,i_p)$ the number of occurrences of $j$ in $(i_1,\dots,i_p)$.

\begin{Definition}\label{def:identity}
    \em The {\em symmetric tensor identity}, denoted $\mathcal{I}^{(p)}_N$ or just $\mathcal{I}$, is the tensor $\mathcal{I}=0$ if $p$ odd and
\[ \mathcal{I}_{i_1,\dots,i_p} = \mathds{1}_{(i_1,\dots,i_p) \in \mathfrak{P}} / \Gamma_{i_1,\dots,i_p} \qquad \mbox{if } p \mbox{ even.}\]
\end{Definition}

{\bf Hermitian tensors.}
Now, let $p$ be an even integer. Let $H \in \bigl(\mathbb{C}^N\bigr)^{\otimes p}$ be a complex tensor of order $p$ and dimension $N$. We say that $H$ is {\em hermitian} if
\smash{$ H = H^{(0)} + i H^{(1)}$},
with~\smash{$H^{(0)} \hspace{-0.5pt} \in\hspace{-0.5pt}  \mathcal{S}^{(p)}(N)$} and \smash{$H^{(1)}\hspace{-1pt} \in\hspace{-0.5pt} \mathcal{A}^{(p)}(N)$}. Hence, for any $(i_1,\dots,i_p)\hspace{-0.5pt} \in \hspace{-0.5pt} \mathfrak{P}$, we have in particular \smash{$H_{i_1,\dots,i_p}=H^{(0)}_{i_1,\dots,i_p} \hspace{-1pt}\in\hspace{-0.5pt}  \mathbb{R}$}. We denote \smash{$\mathcal{H}^{(p)}(N)$} the set of Hermitian tensors of order $p$ and dimension~$N$.

{\bf Self-dual hermitian tensors.}
Let $p \equiv 2 \pmod 4$ be an integer and let $H \in \bigl(\mathbb{C}^{2N}\bigr)^{\otimes p}$ be a~complex tensor of order $p$ and dimension $2N$. The quaternion elements, denoted
\begin{gather*}
 \mathbf{1}=\mathbf{e_0}=\begin{pmatrix} 1 & 0 \\ 0 & 1 \end{pmatrix},\qquad
 \mathbf{e_1}=\begin{pmatrix} {\rm i} & 0 \\ 0 & -{\rm i} \end{pmatrix},\qquad
 \mathbf{e_2}=\begin{pmatrix} 0 & -1 \\ 1 & 0 \end{pmatrix}, \qquad
 \mathbf{e_3}=\begin{pmatrix} 0 & -{\rm i} \\ -{\rm i} & 0 \end{pmatrix},
 \end{gather*}
form a complete set and any $2 \times 2$ complex matrix can be written
\[ \begin{pmatrix} a & b \\ c & d \end{pmatrix}= \frac{a+d}{2}\mathbf{1} -{\rm i}\frac{a-d}{2} \mathbf{e_1} - \frac{b-c}{2}\mathbf{e_2} + {\rm i}\frac{b+c}{2}\mathbf{e_3}.\]
Thus, identifying \smash{$\bigl(\mathbb{C}^{2N}\bigr)^{\otimes p}$} with \smash{$\bigl(\mathbb{C}^{2}\otimes \mathbb{C}^{2}\bigr)^{\otimes p/2}\times \bigl(\mathbb{C}^{N}\bigr)^{\otimes p}$}, the complex tensor $H$ may be canonically written as a tensor $Q$ of order $p$, dimension $N$ and with entries in \smash{$\bigl(\mathbb{C}^{2}\otimes \mathbb{C}^{2}\bigr)^{\otimes p/2}$}, given~by
\[ Q=\sum_{\epsilon_1,\dots,\epsilon_{p/2} \in \{ 0,1,2,3 \}} Q^{(\epsilon_1,\dots,\epsilon_{p/2})} \bigotimes_{s=1}^{p/2} \mathbf{e_{\epsilon_s}}, \]
with \smash{$Q^{(\epsilon_1,\dots,\epsilon_{p/2})} \in \bigl(\mathbb{C}^{N}\bigr)^{\otimes p}$}. Denote $\epsilon=(\epsilon_1,\dots,\epsilon_{p/2})$ and for $0\leq k\leq 3$, $n_k(\epsilon):=\vert \{ s\colon \epsilon_s =k \}\vert$. Let $\mathcal{E}$ be the set of tuples $\epsilon$ such that $n_1(\epsilon)+n_3(\epsilon) \equiv n_2(\epsilon) \pmod 2$, where the number of imaginary and the number of antisymmetric quaternions have same parity.
We say that $Q$ (and then $H$) is {\em self-dual hermitian} if
\[ Q^{(\epsilon)} \in \mathcal{S}^{(p)}(N) \quad \text{if } \epsilon \in \mathcal{E}  \AND Q^{(\epsilon)} \in \mathcal{A}^{(p)}(N) \quad \text{otherwise}.\]
We finally denote $\mathcal{Q}^{(p)}(2N)$ the set of self-dual hermitian tensors of order $p$ and dimension $2N$.

We briefly discuss the motivation from the perspective of theoretical physics. Rephrasing Dyson and Mehta, self-dual Hermitian operators correspond to Hamiltonians of systems (with odd spin) that are invariant under time reversal but not rotationally invariant. More precisely, a quaternion element
\[ q = q^{(0)} \mathbf{1} + q^{(1)} \mathbf{e_1}+ q^{(2)}\mathbf{e_2} + q^{(3)}\mathbf{e_3} \]
is called real if $q^{(0)},q^{(1)},q^{(2)},q^{(3)} \in \mathbb{R}$ \big(note that this differs from $q \in \mathbb{R}^{2 \times 2}$\big). If the quaternion is complex, its {\em quaternion conjugate} is
\[ \bar q = q^{(0)} \mathbf{1} - q^{(1)} \mathbf{e_1} - q^{(2)}\mathbf{e_2} - q^{(3)}\mathbf{e_3}, \]
and its {\em Hermitian conjugate} is
\[ q^\dagger = q^{(0)*} \mathbf{1} - q^{(1)*} \mathbf{e_1} - q^{(2)*}\mathbf{e_2} - q^{(3)*}\mathbf{e_3} . \]
For $H$ a matrix of dimension $2N$ written as $Q=(q_{ij})$ of dimension $N$ with quaternionic entries, time reversal invariance and Hermiticity require respectively
$\bar q_{i j} = q_{ji}$ and \smash{$q_{i j}^\dagger = q_{ji} $},
which gives~\smash{$Q^{(0)}:=\bigl(q_{ij}^{(0)}\bigr)_{i,j}$} is real symmetric and $Q^{(1)}$, $Q^{(2)}$, $Q^{(3)}$ are real antisymmetric. We observe here that the time invariance forces $Q^{(2)}$ to be antisymmetric because $\mathbf{e_2}$ is the only antisymmetric quaternion basis element and the Hermiticity forces $Q^{(1)}$, $Q^{(3)}$ to be antisymmetric because~$\mathbf{e_1}$ and $\mathbf{e_3}$ are the complex quaternion basis elements.
This is the first motivation of our definition of self-dual hermitian tensors. For a tensor ($p$-spin) model, we require $Q^{(\epsilon)}$ to be real symmetric or antisymmetric according to whether the number of antisymmetric quaternions factors~$\mathbf{e_2}$ appearing in the decomposition \smash{$\bigotimes_{s=1}^{p/2} \mathbf{e_{\epsilon_s}}$} has the same parity as the number of complex quaternions factors $\mathbf{e_1}$ and $\mathbf{e_3}$, or not.

\begin{Example}
For $p=2$, this coincides with the classical definition of self-dual Hermitian matrix. In this case,
\[ Q= Q^{(0)} \otimes \mathbf{1} + Q^{(1)} \otimes \mathbf{e_{1}} + Q^{(2)} \otimes \mathbf{e_{2}} + Q^{(3)} \otimes \mathbf{e_{3}}, \]
with \smash{$Q^{(0)}=\bigl( Q^{(0)} \bigr)^{\mathsf T} $} and \smash{$Q^{(k)}=-\bigl( Q^{(k)} \bigr)^{\mathsf T} $} for $k \in \{1,2,3 \}$.

The first genuine tensor example is $p=6$. Here we have $p/2=3$ and for $\epsilon=(\epsilon_1,\epsilon_2,\epsilon_3)$, the parity condition is $n_1(\epsilon)+n_3(\epsilon) \equiv n_2(\epsilon) \pmod 2$, or in other words $\epsilon \in \mathcal{E}$ if the number of non-zero coefficients is odd. We have, for example,
\[ (0,0,0),(1,1,0),(1,3,0) \in \mathcal{E} ,\qquad  \text{but } (1,0,0),(1,1,1),(1,3,2) \notin \mathcal{E}, \]
meaning that $Q^{(0,0,0)}$, $Q^{(1,1,0)}$, $Q^{(1,3,0)}$ are symmetric tensors of order $3$ and $Q^{(1,0,0)}$, $Q^{(1,1,1)}$, $Q^{(1,3,2)}$ are antisymmetric ones.
\end{Example}

{\bf Similarity transformations.}
Let $U$ be a nonsingular $N \times N$ matrix. For $p\geq 1$ and~${N \geq 1}$, the multilinear transformation of a tensor \smash{$H \in \bigl(\mathbb{C}^{N}\bigr)^{\otimes p}$} under $U$ is a tensor denoted $U \cdot H \in \bigl(\mathbb{C}^{N}\bigr)^{\otimes p}$.
Three particular similarity transformations will capture our attention: the cases where~$U$ belongs to $\mathrm{O}(N)$ the orthogonal group $\bigl(U^{\mathsf T}U=I_N\bigr)$, to $\mathrm{U}(N)$ the unitary group \big(${U^*U=I_N}$\big) and to $\mathrm{Sp}(2N)$ the symplectic group \big(${U^{\mathsf T}JU=J}$ with $J=\mathbf{e_2}I^{\mathbb{H}}_N$ in symplectic notation\big). In these cases, the similarity transformations are given by
\begin{itemize}\itemsep=0pt
    \item[--] for $U \in \mathrm{O}(N)$ and $H \in \mathcal{S}^{(p)}(N)$, \[(U \cdot H)_{i_1,\dots,i_p}= \sum_{j_1,\dots,j_p} H_{j_1,\dots,j_p} \prod_{t=1}^p U_{j_t i_t},\]
    \item[--] for $U \in \mathrm{U}(N)$ and $H \in \mathcal{H}^{(p)}(N)$, \[(U \cdot H)_{i_1,\dots,i_p}=\sum_{j_1,\dots,j_p} H_{j_1,\dots,j_p} \prod_{t=1}^{p/2} U_{j_{2t-1} i_{2t-1}} \prod_{t=1}^{p/2} \bar U_{j_{2t} i_{2t}},\]
    \item[--] for $U \in \mathrm{Sp}(2N)$ and $H \in \mathcal{Q}^{(p)}(2N)$, \[(U \cdot H)_{i_1,\dots,i_p}=\sum_{j_1,\dots,j_p} H_{j_1,\dots,j_p} \prod_{t=1}^{p/2} U_{j_{2t-1} i_{2t-1}} \prod_{t=1}^{p/2} (-JUJ)_{j_{2t} i_{2t}}.\]
\end{itemize}

\begin{Definition}\label{def:tens_inv}
    \em A real symmetric $($resp.\ hermitian, resp.\ self-dual hermitian$)$ random tensor~$H$ is said {\em orthogonal invariant} $($resp.\ {\em unitary invariant}, resp.\ {\em symplectic invariant}$)$ if $H$ has the same law as $U \cdot H$ for any $U \in \mathrm{O}(N)$ $($resp.\ $\mathrm{U}(N)$, resp.\ $\mathrm{Sp}(2N))$.
\end{Definition}

\begin{Remark}
    In the complex case, one could also consider the transformation
    \[
    \bigl(H \cdot \bigl(U^{(1)},\dots,U^{(p/2)} \bigr)\bigr)_{i_1,\dots,i_p}=\sum_{j_1,\dots,j_p} H_{j_1,\dots,j_p} \prod_{t=1}^{p/2} U^{(t)}_{j_{2t-1} i_{2t-1}} \prod_{t=1}^{p/2} \bar U^{(t)}_{j_{2t} i_{2t}},
     \]
     where we act with a different unitary on each leg of the tensor. In this case we speak of local unitary invariance and many applications exist in quantum physics. We refer the reader to the rich works of Gurau \cite{guraubook}, Lionni \cite{lionnispringer} or Sasakura \cite{sasakurasigned} for instance.
\end{Remark}

\begin{Remark}
    In the symplectic case, it is also possible to consider the similarity transformation~${(U \cdot H)_{i_1,\dots,i_p}=\sum_{j_1,\dots,j_p} H_{j_1,\dots,j_p} \prod_{t=1}^p U_{j_t i_t}}$, and in this case the trace invariants (see Section \ref{trace_inv}) are multigraphs where edges are decorated by a $J$ instead of multigraphs where odd halfedges are matched with even ones. We do this choice because we want to recover similarity transformations of the type $H \mapsto U^{-1} H U$ in the matrix case.
\end{Remark}

\subsection{Gaussian tensor ensembles}\label{GTE}

{\bf Gaussian orthogonal tensor ensemble.}
A real symmetric tensor $H \in \mathcal{S}_p(N)$ belongs to the~$\textrm{GOTE}(0,1)$ if, as a tensor-valued random variable, it has a density with respect to the natural Lebesgue measure on $\mathcal{S}_p(N)$ proportional to
\[f(H) \propto \exp{\left(- \frac{\| H \| _{\mathrm{F}}^2}{2p}\right)}.\]
The law of such a tensor is orthogonally invariant because the density $f$ only depends on the Frobenius norm, and for all $U \in \mathrm{O}(N)$, $\| U \cdot H \|_{\mathrm{F}} = \|H\|_{\mathrm{F}}$. Moreover, taking into account the symmetry, this also implies that
\smash{$H_{i_1,\dots,i_p} \sim \mathcal{N}\big(0,\frac{p}{ \Gamma_{i_1,\dots,i_p} }\big)$}.
Note that one could also write the variance as
\[ \sigma_{i_1,\dots,i_p}^2:= \frac{p}{ \Gamma_{i_1,\dots,i_p} } = \frac{1}{(p-1)!}\prod_{j=1}^N c_j(i_1,\dots,i_p)!,\]
with again $c_j(i_1,\dots,i_p)$ the number of occurrences of $j$ in $(i_1,\dots,i_p)$. More generally, we say that $H \in \mathcal{S}_p(N)$ belongs to the $\textrm{GOTE}(\beta,\gamma)$ if the density is proportional to
\[f(H) \propto \exp{\left(-\frac{\| H - \beta \mathcal{I} \|_{\mathrm{F}}^2}{2p\gamma}\right)}.\]

{\bf Gaussian unitary tensor ensemble.}
An hermitian tensor $H \in \mathcal{H}_p(N)$ belongs to the $\textrm{GUTE}(0,1)$ if, as a tensor-valued random variable, it has a density with respect to the natural Lebesgue measure on $\mathcal{H}_p(N)$ proportional to
\[f(H) \propto \exp{\left(-\frac{\| H \| _{\mathrm{F}}^2}{p}\right)}.\]
The law of such a tensor is unitary invariant, and the symmetries in the hermitian character of the tensor gives that if $(i_1,\dots,i_p) \in \mathfrak{P}$
\smash{$ H_{i_1,\dots,i_{p}} \sim \mathcal{N}\big(0,\frac{p}{2 \Gamma_{i_1,\dots,i_{p}} }\big)$},
and otherwise, if the indices are not paired,
\[ H_{i_1,\dots,i_{p}} \sim \mathcal{N}\left(0,\frac{p}{2 \Gamma_{i_1,\dots,i_{p}}  }\right) +{\rm i} \mathcal{N}'\left(0,\frac{p}{2 \Gamma_{i_1,\dots,i_{p}} }\right), \]
where $\mathcal{N}$ and $\mathcal{N}'$ are two independent Gaussian laws on $\mathbb{R}$. More generally, we say that ${H \in \mathcal{S}_p(N)}$ belongs to the $\textrm{GUTE}(\beta,\gamma)$ if the density is proportional to
\[f(H) \propto \exp{\left(-\frac{\| H - \beta \mathcal{I} \| _{\mathrm{F}}^2}{p \gamma}\right)}.\]

\begin{Remark}
    One could now define a complex Wigner tensor as \smash{$W=H/N^{\frac{p-1}{2}}$} with $H$ an Hermitian tensor whose entries are centered, have finite moments and the same variance as a~tensor from the GUTE$(0,1)$, and prove the convergence of the tensor moments as done in our previous works for the real case \cite{bonnin2025tensorialfreeconvolutionsemicircular,bonnin2024universalitywignerguraulimitrandom}.
\end{Remark}

{\bf Gaussian symplectic tensor ensemble.}
Finally, a self-dual hermitian tensor $H \in \mathcal{Q}_p(2N)$ belongs to the $\textrm{GSTE}(0,1)$ if, as a tensor-valued random variable, it has a density with respect to the natural Lebesgue measure on $\mathcal{Q}_p(2N)$ proportional to
\[f(H) \propto \exp{\left(-\frac{2 \| H \| _{\mathrm{F}}^2}{p}\right)}.\]
The law of such a tensor is symplectic invariant. Remark that for a self-dual hermitian tensor the entries of all the $Q^{(\epsilon)}$ are in particular real, so the entries $q:=Q_{i_1,\dots,i_p}$ of the tensor~$Q$ satisfy for $s\in \{1,\dots,p/2\}$ and~${\iota=(\iota_1,\dots,\iota_p) \in \{1,2\}^p}$,
$q_{\iota} = \bar q_{\tau_s(\iota)} \text{ if } \iota_{2s-1} = \iota_{2s}$ and~${q_{\iota} = -\bar q_{\tau_s(\iota)}}$ if~${\iota_{2s-1} \neq \iota_{2s}}$,
where $\tau_s\colon (\iota_1,\dots,\iota_p) \mapsto (\iota_1,\dots,3-\iota_{2s-1},3-\iota_{2s},\dots,\iota_p)$ is the partial symmetry. It is similar to what happens in the matrix case where the quaternionic elements, denoted~$\mathbb{H}$, are the ones with real coordinates in the quaternion basis $(e_k)_{0\leq k\leq 3}$, that are matrices of the form~${\bigl(\begin{smallmatrix} z & -\omega \\ \bar \omega & \bar z \end{smallmatrix}\bigr)}$.
The symmetries in the self-dual hermitian character give this time: if~${(i_1,\dots,i_p) \in \mathfrak{P}}$,
\[ q_{1,\dots,1} \sim \mathcal{N}\left(0,\frac{p}{4 \Gamma_{i_1,\dots,i_{p}} }\right) \AND q_{1,\dots,1,2}=0, \]
and otherwise, if the indices are not paired,
\begin{gather*}
 q_{1,\dots,1} \sim \mathcal{N}\left(0,\frac{p}{4 \Gamma_{i_1,\dots,i_{p}} }\right) +{\rm i} \mathcal{N}'\left(0,\frac{p}{4 \Gamma_{i_1,\dots,i_{p}} }\right),  \\ q_{1,\dots,1,2}=\mathcal{N}\left(0,\frac{p}{4 \Gamma_{i_1,\dots,i_{p}} }\right) +{\rm i} \mathcal{N}'\left(0,\frac{p}{4 \Gamma_{i_1,\dots,i_{p}} }\right) ,
 \end{gather*}
where $\mathcal{N}$ and $\mathcal{N}'$ are again independent Gaussian laws on $\mathbb{R}$. More generally, we say that $H \in \mathcal{S}_p(N)$ belongs to the $\textrm{GSTE}(\beta,\gamma)$ if the density is proportional to
\[f(H) \propto \exp{\left(-\frac{2 \| H - \beta \mathcal{I} \| _{\mathrm{F}}^2}{p \gamma}\right)}.\]

\subsection{Trace invariants}\label{trace_inv}

If $H$ and $H'$ are two tensors of order $p$ and $p'$ and same dimension $N$, and $r\leq p,p'$, the {\em contraction} of $H$ and $H'$ along their $r$ first legs, is defined as the $(p+p'-2r)$-order tensor
\[ \bigl(H \star_r H'\bigr)_{i_1,\dots,i_{p+p'-2r}} = \sum_{j_1,\dots,j_{r}}  H_{j_1,\dots,j_r,i_1,\dots,i_{p-r}} H'_{j_1,\dots,j_{r},i_{p-r+1},\dots,i_{p+p'-2r}} . \]
To contract along legs different from the first ones, one could previously act on $H$ and $H'$ by permutations, and in particular, one may get the tensor resulting of the contraction of the $r$-th leg of $H$ with the $t$-th one of $H'$ by $H^{(1,r)} \star_1 H'^{(1,t)}$. The contraction operation is crucial in the tensor setting. In particular, it enables to construct a family that generates all the polynomials of the tensor invariant under orthogonal, unitary or symplectic transformation. This complete family is given by the so-called {\em trace invariants}, see Section \ref{basis} for their detailed definition and the proofs that they form a complete family in each case. They are encoded by $p$-regular multigraph; where the odd halfedges must be matched with even halfedges for the unitary and symplectic cases. More precisely, given a $p$-regular multigraph with $k$ vertices (and $kp/2$ edges), we decorate the vertices by the tensor $H$ and the edges by indices $i_1,\dots,i_{kp/2}$. Then the trace invariant is given by the sum, over each index going from $1$ to the dimension $N$, of the product over the vertices of the entry of $H$ given by the $p$ indices around the vertex. A detailed example is given in Example \ref{example2} in the beginning of Section \ref{basis} where trace invariants are properly introduced.
We call the {\em rank} of a trace invariant the number of vertices in the associated multigraph.

{\bf Rank one and two.}
If $p$ is odd, there is no trace invariant of rank one. Otherwise, if $p$ is even, there is only one trace invariant of rank one which is associated to the {\em bouquet} graph (one vertex and $p/2$ self-loops, see Figure~\ref{fig:melon}). The rank-one trace invariant is given by the {\em paired trace} of $H$ defined as follows:
\[ \operatorname{Tr}^{\bullet \bullet}(H) =
    \begin{cases}
        0 & \mbox{if } p \mbox{ odd,} \\
      \displaystyle  \sum_{i_1,\dots, i_{p/2}} H_{i_1,i_1,\dots,i_{p/2},i_{p/2}} & \text{if } p \mbox{ even.}
    \end{cases}
\]
For the case of rank two, the trace invariants are associated to the graphs with two vertices, a~number $1\leq r \leq p$ of edges between these two vertices with $p-r$ even, and the other edges are self-loops. A distinguished trace invariant of rank two is particularly interesting: the Frobenius norm $\| H \|_{\mathrm{F}}^2$, associated to the {\em melon} multigraph, see Figure~\ref{fig:melon}. It corresponds to the case where the two vertices are linked by $r=p$ edges and there is no self-loop; in the hermitian case the $t$-th halfedge of the first vertex is matched with the $(t+1)$-th one of the second vertex ($p+1=1$) and in the self-dual hermitian case the $2t$-th one of a vertex is matched with the $(2t-1)$-th one of the other vertex. The computation proving that the trace invariant associated to the melon is the Frobenius norm is done in Section \ref{melonfrob}. Note finally that we have
\[\| H -\beta \mathcal{I} \| _{\mathrm{F}}^2 / \gamma = a\| H \| _{\mathrm{F}}^2 +b \operatorname{Tr}^{\bullet \bullet}(H) + c ,\]
with $a=1/ \gamma$, $b= - 2 \beta / \gamma$ and $c= \beta^2 \|\mathcal{I} \| _{\mathrm{F}}^2 / \gamma$.

\begin{figure}[ht]
    \begin{minipage}[c]{.3\linewidth}
        \centering
        \begin{tikzpicture}[scale=0.45]
    \draw[fill=black] (0,0) circle (5pt);
    \draw (0,0) .. controls (3.236,2.351) and (3.236,-2.351) .. (0,0);
    \draw (0,0) .. controls (3.236,2.351) and (0-1.236,3.804) .. (0,0);
    \draw (0,0) .. controls (3.236,-2.351) and (0-1.236,-3.804) .. (0,0);
    \draw (0,0) .. controls (0-1.236,3.804) and (-4,0) .. (0,0);
    \draw (0,0) .. controls (0-1.236,-3.804) and (-4,0) .. (0,0);
\end{tikzpicture}
    \end{minipage}
    \quad
    \begin{minipage}[c]{.3\linewidth}
        \centering
        \begin{tikzpicture}[scale=0.5]
    \draw[fill=black] (0,0) circle (5pt);
    \filldraw[black] (4,0) circle (5pt);
    \draw (0,0) .. controls (0+1.33,2) and (0+2.67,2) .. (4,0);
    \draw (0,0) .. controls (0+1.33,1) and (0+2.67,1) .. (4,0);
    \draw (0,0) -- (4,0) ;
    \draw (0,0) .. controls (0+1.33,-2) and (0+2.67,-2) .. (4,0);
    \draw (0,0) .. controls (0+1.33,-1) and (0+2.67,-1) .. (4,0);
\end{tikzpicture}
    \end{minipage}
    \quad
    \begin{minipage}[c]{.3\linewidth}
        \centering
        \begin{tikzpicture}[scale=0.5]
    \draw[fill=black] (0,0) circle (5pt);
    \filldraw[black] (4,0) circle (5pt);
    \draw (0,0) .. controls (0-2,2) and (0-2,-2) .. (0,0);
    \draw (4,0) .. controls (4+2,2) and (4+2,-2) .. (4,0);
    \draw (0,0) .. controls (0+1.33,1) and (0+2.67,1) .. (4,0);
    \draw (0,0) -- (4,0) ;
    \draw (0,0) .. controls (0+1.33,-1) and (0+2.67,-1) .. (4,0);
\end{tikzpicture}
    \end{minipage}
\caption{Bouquet graph ($p=10$), melon graph ($p=5$), another trace invariant of rank $2$ ($p=5$), respectively $m_G(H)=\operatorname{Tr}^{\bullet \bullet}(H)$, $m_G(H)=\| H \|^2_F$, $m_G(H)=\sum_{a,b,c,d,e} H_{aabcd} H_{ebcde}$.}
\label{fig:melon}
\end{figure}

\begin{Remark}
    In the vector case, there exists only one trace invariant (associated to the melon graph with two vertices and one edge) given by $\| H \|_{\mathrm{F}}^2=\vert H \vert^2_2$. In the matrix case, since the $2$-regular multigraph are exactly the cycles, the trace invariants are the traces of powers of $H$ (it is known since Weyl \cite{wey66} that \smash{$\bigl(\operatorname{Tr}\bigl(H^k\bigr)\bigr)_{k\leq N}$} is a basis of invariants).
\end{Remark}

\subsection{Maxwell theorem}

The original Maxwell theorem states that a random vector of dimension two or more has independent entries and is rotationally invariant if and only it has a law with a density of the form~${H \mapsto \exp\bigl(-a\vert H \vert_2^2 + c\bigr)}$,
with $(a,c) \in \mathbb{R}^*_+ \times \mathbb{R}$. The generalization of this result to the order two gives that a real symmetric (resp.\ hermitian, resp.\ self-dual hermitian) random matrix of dimension two or more has in the same time independent entries (up to symmetries) and a law invariant by orthogonal (resp.\ unitary, resp.\ symplectic) transformation if and only if its law is proportional to the one of the {\em GOE} (resp.\ {\em GUE}, resp.\ {\em GSE}) eventually dilated and translated, that is, a law with a density of the form
\[ H \mapsto \exp \left( - a \operatorname{Tr}\bigl(H^2\bigr) + b \operatorname{Tr}(H) + c \right),\]
where $(a,b,c) \in \mathbb{R}^*_+ \times \mathbb{R} \times \mathbb{R}$. A complete proof can be found in the monograph of Mehta~\cite{Meh2004}, our general proof will adapt this one in the tensor setting. Indeed, with the objects we introduced previously, both vector and matrix result may be unified in an overall characterization of Gaussian tensor ensembles for any $p\geq 1$.

\begin{Theorem}[main]\label{maxwell}
For any $p\geq 1$ and $N\geq 2$, a real symmetric $($resp.\ hermitian, resp.\ self-dual hermitian$)$ random tensor $H$ of order $p$ and dimension $N$ has in the same time independent entries $($up to symmetries$)$ and a law invariant by orthogonal $($resp.\ unitary, resp.\ symplectic$)$ transformation if and only if it belongs to the Gaussian orthogonal $($resp.\ unitary, resp.\ symplectic$)$ tensor ensemble, that is, it has a law with a density of the form
\[ H \mapsto \alpha \exp \left( - \| H - \beta \mathcal{I} \|_{\mathrm{F}}^2 / \gamma \right),\]
where $(\alpha,\beta,\gamma) \in \mathbb{R}^*_+ \times \mathbb{R} \times \mathbb{R}^*_+$,
or equivalently,
\[ H \mapsto \exp \left( - a \| H \|_{\mathrm{F}}^2 + b \operatorname{Tr}^{\bullet \bullet}(H) + c \right),\]
where $(a,b,c) \in \mathbb{R}^*_+ \times \mathbb{R} \times \mathbb{R}$.
\end{Theorem}

It is very insightful to understand what does this result means in terms of invariants. The theorem proves that independence of the entries and invariance force the law to depend in the same way on only two very particular trace invariants: the Frobenius norm (associated to the melon graph) and the paired trace (associated to the bouquet graph, existing only when $p$ is even).

{\bf Letac extension.}
There exists an extension of the original Maxwell theorem due to Le\-tac~\cite{letac} for $N\geq 3$, with counterexamples for $N=2$. He observes that a random vector $H \in \mathbb{R}^N$ has independent entries and is ``isotropic'', that is, $\mathbb{P}(H=0)=0$ and $H/\vert H\vert_2$ is uniformly distributed on the sphere, if and only if its components are independent centered Gaussian variables with the same positive variance. It implies Maxwell theorem when $N\geq 3$ because rotational invariance and independence of the entries imply isotropy for three or more variables, but isotropy contrasts strongly with rotational invariance for two independent random variables. To the best of the author's knowledge, no generalization to matrices or higher orders exists, we discuss about it in Section \ref{sec:let}. We will therefore consider a real symmetric random tensor $H$ with independent entries up to symmetries such that $H / \| H \|_{\mathrm{F}}$ is uniformly distributed on the Frobenius sphere.

\section{Proof of Maxwell theorem}

\subsection{Warm-up: Proof of the original theorem for vectors}\label{vector}

First, the law of a Gaussian vector belonging to \smash{$\mathcal{N}\bigl(0,\sigma^2 I_N\bigr) = \mathcal{N}\bigl(0,\sigma^2\bigr)^{\otimes N}$}, having for density \smash{$H \mapsto \exp\bigl(\frac{1}{2}\vert H \vert_2^2 - \frac{N}{2} \ln\bigl(2 \pi \sigma^2\bigr)\bigr)$}, is product and rotationally invariant. Reciprocally, let $N\geq 2$ and $\mu$ be a law on $\mathbb{R}^N$ product and rotationally invariant. Assume in a first time that $\mu$ has a~smooth density $f\colon \mathbb{R}^N \mapsto (0,\infty)$. As $\mu$ is product, $f$ must be of the form $f(H)=\prod_{i=1}^N f_i(H_i)$, that is,
\[ \ln f(H) = \sum_{i=1}^N \ln f_i(H_i) =: \sum_{i=1}^N g_i(H_i).\]
Hence, taking the derivative along one coordinate we get
$\partial_i \ln f(H) = g'_i(H_i)$,
so $\partial_i \ln f(H)$ must depend only on $H_i$. Moreover, the rotational invariance implies $\ln f(H)=\psi \bigl(\vert H \vert^2_2\bigr)$, so we have
\[ \partial_i \ln f(H) = 2 \psi'\bigl(\vert H \vert^2_2\bigr) H_i.\]
As $\partial_i \ln f(H)$ depends only on $H_i$ and $N \geq 2$, then $\psi'$ must be constant. That gives that there exists $a$, $b$ such that $\psi(x)=ax+b$, so finally $f(H)=e^{a \vert H \vert_2^2+b}$ and $a<0$ because $f$ is a density.

It remains to prove that we can assume without loss of generality that $\mu$ has a smooth density. Otherwise, take $H_{\epsilon}=H+\epsilon W$ where $W \sim \mathcal{N}(0,I_N)$. It has a smooth density given by $\mu * \mathcal{N}(0,I_N)$, it is still rotationally invariant and has independent entries, so it is Gaussian according to what we just proved. Hence, $H_{\epsilon}$ and $W$ are $L^2$ and centered so $H$ too. The two first moments of $H_{\epsilon}$ converge to the ones of $H$ and the Fourier transform of $H_{\epsilon}$ satisfies~${\phi_{H_{\epsilon}}(t)=\phi_H(t)\phi_{W}(\epsilon t) \rightarrow \phi_H(t)}$ when $\epsilon$ goes to zero. As a consequence, $H_{\epsilon}$ being Gaussian, we obtain that $H$ is Gaussian too.

\subsection{Proof of the main theorem}

We first consider the orthogonal case, we will give the details of the slight modifications necessary to obtain the two others ones in a second time. In order to prove the main theorem, let us introduce, for $\theta \in [0,2\pi]$,
\[U_{\theta}=\begin{pmatrix}
    \cos{\theta} & \sin{\theta} & 0 & \cdots & 0 \\
    -\sin{\theta} & \cos\theta & 0 & \cdots & 0 \\
    0 & \cdots & 1 & \cdots & 0 \\
    \cdots & \cdots & \cdots & \ddots & \cdots \\
    0 & \cdots & 0 & \cdots & 1 \\
\end{pmatrix}\]
and denote
\[ A= \frac{\partial U_{\theta}^{\mathsf T}}{\partial \theta} U_{\theta} = \begin{pmatrix}
    0 & -1 &  \cdots & 0 \\
    1 & 0 &  \cdots & 0 \\
    \cdots & \cdots & \ddots & \cdots \\
    0 & \cdots  & \cdots & 0 \\
\end{pmatrix}.\]
The proof will rely on the fact that if $H$ has a law orthogonally invariant, then the law of $H$ and $H'=U_{\theta} \cdot H$ is the same and one may differentiate with respect to $\theta$. We first state some useful intermediate results that we are going to prove in a second time.
\begin{Proposition}\label{derivatives}
We have
    \[ \left( \frac{\partial H}{\partial \theta} \right) = \sum_{r=1}^p A \star_1 H^{(1,r)}. \]
In other words,
\[\left( \frac{\partial H}{\partial \theta} \right)_{i_1,\dots,i_p} = \sum_{k=1}^N (A_{i_1k}H_{k,i_2,\dots,i_p} + A_{i_2k}H_{i_1,k,\dots,i_p} + \dots + A_{i_pk}H_{i_1,\dots,i_{p-1},k}).\]
\end{Proposition}

\begin{Lemma}\label{xy}
Let $g_1$, $g_2$, $g_3$ be three continuous and differentiable functions such that
$ g_1(xy)=g_2(x)+g_3(y)$.
Then, there exists $\alpha$, $\beta_1$, $\beta_2$, $\beta_3$ such that for all $k\in \{1,2,3\}$,
$g_k(u)=\alpha \ln u + \beta_k$,
with $\beta_1 = \beta_2+\beta_3$.
\end{Lemma}

\begin{proof}[Proof of Theorem~\ref{maxwell}.]
Let $H$ be a real symmetric tensor having independent entries and a~law $\mu$ orthogonally invariant. Possibly after changing $H$ into $H+\epsilon W$ with $W$ belonging to the GOTE, we may assume without loss of generality that $\mu$ has a smooth density given by
\[ f(H)=\prod_{i_1\leq \dots \leq i_p} f_{i_1,\dots,i_p}(H_{i_1,\dots,i_p}). \]
By orthogonal invariance,
$\frac{\partial}{\partial \theta} \ln f(H)=0$.
Then using the computation of Proposition \ref{derivatives}, we find
\begin{align*}
    0 &= \sum_{i_1\leq \dots \leq i_p} \frac{1}{f_{i_1,\dots,i_p}(H_{i_1,\dots,i_p})} \frac{\partial f_{i_1,\dots,i_p}}{\partial H_{i_1,\dots,i_p}} \frac{\partial H_{i_1,\dots,i_p}}{\partial \theta} \\
    &= \sum_{i_1\leq \dots \leq i_p} \frac{1}{f_{i_1,\dots,i_p}} \frac{\partial f_{i_1,\dots,i_p}}{\partial H_{i_1,\dots,i_p}} \sum_{k=1}^N (A_{i_1k}H_{k,i_2,\dots,i_p} + A_{i_2k}H_{i_1,k,\dots,i_p} + \dots + A_{i_pk}H_{i_1,\dots,i_{p-1},k}).
\end{align*}
Moreover, we know that $A_{i,k}$ is equal to $-1$ if $(i,k)=(1,2)$, to $1$ if $(i,k)=(2,1)$ and to $0$ otherwise. We introduce a brief notation for ease of notation in the sums with multi-indices. For given integers $r \in \{1,\dots, p-1\}$ and $i_{r+1},\dots,i_p$, let us denote $f_{a_1,\dots,a_r}$ for $f_{a_1,\dots,a_r,i_{p-r},\dots,i_p}$ and $H_{a_1,\dots,a_r}$ for $H_{a_1,\dots,a_r,i_{p-r},\dots,i_p}$. In particular, $f_1(H_1)=f_{1,i_{2},\dots,i_p}(H_{1,i_{2},\dots,i_p})$. The previous equations gives finally
\begin{align}
    0= {}&\sum_{3\leq i_2\leq \dots \leq i_p} \left[ \frac{-1}{f_1(H_1)} \frac{\partial f_{1}}{\partial H_{1}} H_2 + \frac{1}{f_2(H_2)} \frac{\partial f_{2}}{\partial H_{2}} H_1 \right]\nonumber\\
    &+ \sum_{3\leq i_3\leq \dots \leq i_p} \left[ 2 \left( \frac{-1}{f_{1,1}(H_{1,1})} \frac{\partial f_{1,1}}{\partial H_{1,1}} + \frac{1}{f_{2,2}(H_{2,2})} \frac{\partial f_{2,2}}{\partial H_{2,2}} \right) H_{1,2}\right.\nonumber\\
    & \left.+ \frac{1}{f_{1,2}(H_{1,2})} \frac{\partial f_{1,2}}{\partial H_{1,2}} (H_{1,1} - H_{2,2}) \right]+ \sum_{3\leq i_4\leq \dots \leq i_p} \left[ \dots \right].\label{eq-A}
\end{align}
The different brackets depend on mutually exclusive sets of variables, and their sum is $0$, so each must be a constant and in particular for given integers $i_p\geq \dots \geq i_2 \geq 3$, there exists~${C=C_{i_2,\dots,i_p}}$ such that
\[C = \frac{-1}{f_1(H_1)} \frac{\partial f_{1}}{\partial H_{1}} H_2 + \frac{1}{f_2(H_2)} \frac{\partial f_{2}}{\partial H_{2}} H_1. \]
Hence, after dividing by $H_1H_2$ we get
\[ \frac{C}{H_1H_2} = \frac{-1}{H_1} \frac{1}{f_1(H_1)} \frac{\partial f_{1}}{\partial H_{1}} + \frac{1}{H_2}\frac{1}{f_2(H_2)} \frac{\partial f_{2}}{\partial H_{2}}, \]
and now applying Lemma \ref{xy} we find that the constant $C$ must be zero. That means
\[ \frac{1}{H_1} \frac{1}{f_1(H_1)} \frac{\partial f_{1}}{\partial H_{1}} = \frac{1}{H_2}\frac{1}{f_2(H_2)} \frac{\partial f_{2}}{\partial H_{2}}, \]
and this must be a constant $c'=c'_{i_2,\dots,i_p}$, which gives
\[\frac{1}{f_1(H_1)} \frac{\partial f_{1}}{\partial H_{1}} = c' H_1.\]
After integrating we find $\ln f_1(H_1) = \frac{c'}{2} H_1^2$, so $c'$ must be negative ($f_1$ is a density) and we denote it $-2a$ in place of, with $a \in \mathbb{R}^*_+$ (depending on ${i_2,\dots,i_p}$). Finally, we conclude from all of this
\smash{$f_{1,i_{2},\dots,i_p}(H_{1,i_{2},\dots,i_p}) = \exp\bigl(-a H_{1,i_{2},\dots,i_p}^2\bigr)$}.
The same holds for $f_2(H_2)$ giving $f_{2,i_{2},\dots,i_p}(H_{2,i_{2},\dots,i_p}) = \exp\bigl(-a H_{2,i_{2},\dots,i_p}^2\bigr)$.
Note that involving the symmetries and orthogonal matrices $U_{\theta}$ with the rotation acting at different positions than~$(1,2)$, it follows that any $p$-tuple of indices with at least {\em one index $i_k$ distinct from all other ones} gives a contribution~${f_{i_{1},\dots,i_p}(H_{i_{1},\dots,i_p}) = \exp\bigl(-a_{i_1,\dots,i_{k-1},i_{k+1},\dots,i_p} H_{i_{1},\dots,i_p}^2\bigr)}$. Now, the second bracket in equation~\eqref{eq-A} gives for $i_p\geq\dots\geq i_3\geq 3$,
\[ 2 \left( \frac{-1}{f_{1,1}(H_{1,1})} \frac{\partial f_{1,1}}{\partial H_{1,1}} + \frac{1}{f_{2,2}(H_{2,2})} \frac{\partial f_{2,2}}{\partial H_{2,2}} \right) -2a_{1,2,i_3,\dots,i_p} (H_{1,1} - H_{2,2}) = \frac{C_{i_3,\dots,i_p}}{H_{1,2}} =0,\]
and we deduce immediately
\[ \frac{1}{f_{1,1}(H_{1,1})} \frac{\partial f_{1,1}}{\partial H_{1,1}} + a H_{1,1} = \frac{1}{f_{2,2}(H_{2,2})} \frac{\partial f_{2,2}}{\partial H_{2,2}} + a H_{2,2} = b_{i_3,\dots,i_p},\]
that is, finally
\[ f_{1,1,i_{3},\dots,i_p}(H_{1,1,i_{3},\dots,i_p}) = \exp\left(-\frac{a}{2} H_{1,1,i_{3},\dots,i_p}^2-bH_{1,1,i_{3},\dots,i_p}\right). \]
But as all invariants are expressible in terms of the trace invariants, the two only ones which can appear are the melon (corresponding to the Frobenius norm) and the bouquet (corresponding to the paired trace), and the constants must satisfy \smash{$a_{i_1,\dots,i_p}=\frac{a_{1,\dots,p}}{\Gamma_{i_1,\dots,i_p} }$}, $b_{i_1,i_1,\dots,i_{p/2},i_{p/2}}=\smash{\frac{b}{\Gamma_{i_1,i_1,\dots,i_{p/2},i_{p/2}} }}$ and $b_{i_1,\dots,i_p}=0$ otherwise, leading to
\begin{align*}
    f(H)={}& \exp \Bigg( -\sum_{i_1\leq\dots\leq i_p} \frac{a}{\Gamma_{i_1,\dots,i_p} } H_{i_{1},\dots,i_p}^2 \\
    &+ \mathds{1}_{p \text{ even}} \sum_{i_1 \leq\dots \leq i_{p/2}} \frac{b}{\Gamma_{i_1,i_1,\dots,i_{p/2},i_{p/2}} } H_{i_1,i_1,\dots,i_{p/2},i_{p/2}} +c \Bigg) \\
    ={}& \exp \Bigg( -a \sum_{i_1,\dots,i_p} H_{i_{1},\dots,i_p}^2 + b \mathds{1}_{p \text{ even}} \sum_{i_1,\dots, i_{p/2}} H_{i_1,i_1,\dots,i_{p/2},i_{p/2}} +c \Bigg).
\end{align*}
This concludes the proof.
\end{proof}

\begin{proof}[Proof of Proposition \ref{derivatives}.]
Consider the transformation $H=U \cdot H'$ with $U=U_{\theta}$. Then we have the equality
\begin{align*}
    \sum_{k=1}^N U_{i_1k} H_{k,i_2,\dots,i_p} &= \sum_{j_1,\dots,j_p} H'_{j_1,\dots,j_p} \prod_{t=2}^p U_{j_ti_t} \sum_{k=1}^N U_{i_1k}U^{\mathsf T}_{k,j_1} \\
    &= \sum_{j_1,\dots,j_p} H'_{j_1,\dots,j_p} \prod_{t=2}^p U_{j_ti_t} \times \mathbf{1}_{j_1=i_1},
\end{align*}
which finally gives
\begin{equation}\label{HH'}
    \sum_{k=1}^N U_{i_1k} H_{k,i_2,\dots,i_p} = \sum_{j_2,\dots,j_p} H'_{i_1,j_2,\dots,j_p} \prod_{t=2}^p U_{j_ti_t}
\end{equation}
Now one can compute the differentiation of $H$ with respect to $\theta$,
\begin{align*}
    \left( \frac{\partial H}{\partial \theta} \right)_{i_1,\dots,i_p} &= \sum_{j_1,\dots,j_p} H'_{j_1,\dots,j_p} \sum_{r=1}^p \prod_{\genfrac{}{}{0pt}{}{t=1}{t\neq r}  }^p U_{j_ti_t} \left( \frac{\partial U_{j_r,i_r}}{\partial \theta} \right) \\
    &= \sum_{r=1}^p \sum_{j_r} \left( \frac{\partial U_{j_r,i_r}}{\partial \theta} \right) \sum_{ \genfrac{}{}{0pt}{}{j_1,\dots,j_p}{ \widehat{j_r}} } H'_{j_1,\dots,j_p} \prod_{\genfrac{}{}{0pt}{}{t=1}{t\neq r} }^p U_{j_ti_t} \\
    &= \sum_{r=1}^p \sum_{k=1}^N \left( \sum_{j_r} \left( \frac{\partial U}{\partial \theta} \right)^{\mathsf T}_{i_r,j_r} \times U_{j_rk} \right) H_{i_1,\dots,i_{r-1},k,i_{r+1},\dots,i_p},
\end{align*}
where we used equation~\eqref{HH'} for the last line. Replacing by $A$, we finally get the result
\[\left( \frac{\partial H}{\partial \theta} \right)_{i_1,\dots,i_p} = \sum_{k=1}^N (A_{i_1k}H_{k,i_2,\dots,i_p} + A_{i_2k}H_{i_1,k,\dots,i_p} + \dots + A_{i_pk}H_{i_1,\dots,i_{p-1},k}).\tag*{\qed}\]\renewcommand{\qed}{}
\end{proof}

\begin{proof}[Proof of Lemma \ref{xy}]
Differentiating the relation $g_1(xy)=g_2(x)+g_3(y)$ with respect to $x$ gives
$ g'_1(xy)=yg'_2(x)$,
and after integrating with respect to $y$, we get
$g_1(xy)=xg'_2(x)\ln y + h(x)$
Hence, inserting this equation in the initial relation, we have
\[ g_3(y) = g_1(xy)-g_2(x) = xg'_2(x)\ln y + h(x) - g_2(x).\]
The right-hand side term must be independent of $y$ which means that there exists constants~$a$,~$b_3$ such that $xg'_2(x)=a$ and $h(x) - g_2(x)=b_3$, that is,
$ g_3(y)=a \ln y + b_3$,
and by integration of~${g'_2(x)=a/x}$ we also have
$ g_2(x)=a \ln x + b_2$.
Finally, this gives $g_1(xy)=a \ln(xy) + b_2 + b_3$.
\end{proof}

{\bf Unitary and symplectic cases.}
We adapt the proof for hermitian and self-dual hermitian tensors. For $\theta \in [0,2\pi]$, we take the same matrix $U_{\theta}$ which is both orthogonal, unitary and symplectic (of size $2N$ in this last case), that is,
\[U_{\theta}=\begin{pmatrix}
    \cos{\theta} & \sin{\theta} & 0 & \cdots & 0 \\
    -\sin{\theta} & \cos\theta & 0 & \cdots & 0 \\
    0 & \cdots & 1 & \cdots & 0 \\
    0 & \cdots & \cdots & \ddots & 0 \\
    0 & \cdots & 0 & \cdots & 1 \\
\end{pmatrix}
\AND
A= \frac{\partial U_{\theta}^{\mathsf T}}{\partial \theta} U_{\theta} = \begin{pmatrix}
    0 & -1 &  \cdots & 0 \\
    1 & 0 &  \cdots & 0 \\
    0 & \cdots & \ddots & 0 \\
    0 & \cdots  & \cdots & 0 \\
\end{pmatrix},\]
or in quaternionic notations
\[U_{\theta}=\begin{pmatrix}
    \mathbf{1} \cos{\theta} - \mathbf{e_2} \sin{\theta} & 0 & \cdots & 0 \\
    0 & \mathbf{1} & \cdots & 0 \\
    \cdots & \cdots & \ddots & \cdots \\
    0 & \cdots &  \cdots & \mathbf{1} \\
\end{pmatrix}
\AND
A = \begin{pmatrix}
    \mathbf{e_2} & 0 &  \cdots & 0 \\
    0 & 0 &  \cdots & 0 \\
    \cdots & \cdots & \cdots & \cdots \\
    0 & \cdots  & \cdots & 0 \\
\end{pmatrix}.\]
The computation of the derivatives of $H$ with respect to $\theta$ is exactly the same. Now recall that if~$H$ is hermitian (resp.\ $\mathrm{H}$ is self-dual hermitian identified as $Q$ with $(p/2)$-quaternionic entries), it is of the form \smash{$H = H^{(0)}+{\rm i}H^{(1)}$} \big(resp.\ \smash{$Q=\sum_{\epsilon \in \mathcal{E} } Q^{(\epsilon)} \bigotimes_{s=1}^{p/2} \mathbf{e_{\epsilon_s}} $}\big) with \smash{$H^{(0)} \in \mathcal{S}^{(p)}(N)$}, \smash{$H^{(1)} \in \mathcal{A}^{(p)}(N)$} \big(resp.\ $Q^{(\epsilon)} \in \mathcal{S}^{(p)}(N)$ if $\epsilon \in \mathcal{E}$, $Q^{(\epsilon)} \in \mathcal{A}^{(p)}(N)$ otherwise\big). Hence if $H$, resp.~$\mathrm{H}$, is a tensor hermitian, resp.\ self-dual hermitian, with independent entries and a~law unitary, resp.\ symplectic, invariant, their laws have densities (without loss of generality after adding a~tensor from the GUTE, resp.\ GSTE) of the form
\[ f(H)= \prod_{i_1\leq \dots \leq i_p} f_{i_1,\dots,i_p}^{(0)}\bigl(H^{(0)}_{i_1,\dots,i_p}\bigr) \prod_{ \genfrac{}{}{0pt}{}{i_1\leq \dots \leq i_p }{\text{not paired}} } f_{i_1,\dots,i_p}^{(1)}\bigl(H^{(1)}_{i_1,\dots,i_p}\bigr), \]
and
\[ f(\mathrm{H})= \prod_{\epsilon \in \mathcal{E}} \prod_{i_1\leq \dots \leq i_p} f_{i_1,\dots,i_p}^{(\epsilon)}\bigl(Q^{(\epsilon)}_{i_1,\dots,i_p}\bigr) \prod_{\epsilon \notin \mathcal{E}} \prod_{\genfrac{}{}{0pt}{}{i_1\leq \dots \leq i_p }{\text{not paired}} } f_{i_1,\dots,i_p}^{(\epsilon)}\bigl(Q^{(\epsilon)}_{i_1,\dots,i_p}\bigr). \]
Denote $\mathbf{i}=(i_1,\dots,i_p)$. Then, the invariance under respective similarity transformations gives
\begin{align*}
    0 &= \frac{\partial}{\partial \theta} \ln f(H) \\
    &= \left[\sum_{i_1\leq \dots \leq i_p} \frac{1}{f^{(0)}_{\mathbf{i}}(H^{(0)}_{\mathbf{i}})} \frac{\partial f^{(0)}_{\mathbf{i}}}{\partial H^{(0)}_{\mathbf{i}}} \frac{\partial H^{(0)}_{\mathbf{i}}}{\partial \theta} \right] + \left[ \sum_{\genfrac{}{}{0pt}{}{i_1\leq \dots \leq i_p }{\text{not paired}} } \frac{1}{f^{(1)}_{\mathbf{i}}(H^{(1)}_{\mathbf{i}})} \frac{\partial f^{(1)}_{\mathbf{i}}}{\partial H^{(1)}_{\mathbf{i}}} \frac{\partial H^{(1)}_{\mathbf{i}}}{\partial \theta} \right] ,
\end{align*}
and
\[ 0= \sum_{\epsilon \in \mathcal{E}} \left[ \sum_{i_1\leq \dots \leq i_p} \frac{1}{f^{(\epsilon)}_{\mathbf{i}}(Q^{(\epsilon)}_{\mathbf{i}})} \frac{\partial f^{(\epsilon)}_{\mathbf{i}}}{\partial Q^{(\epsilon)}_{\mathbf{i}}} \frac{\partial Q^{(\epsilon)}_{\mathbf{i}}}{\partial \theta} \right] + \sum_{\epsilon \notin \mathcal{E}} \left[ \sum_{\genfrac{}{}{0pt}{}{i_1\leq \dots \leq i_p }{\text{not paired}} } \frac{1}{f^{(\epsilon)}_{\mathbf{i}}(Q^{(\epsilon)}_{\mathbf{i}})} \frac{\partial f^{(\epsilon)}_{\mathbf{i}}}{\partial Q^{(\epsilon)}_{\mathbf{i}}} \frac{\partial Q^{(\epsilon)}_{\mathbf{i}}}{\partial \theta} \right].\]
Each bracket in both equations depends on mutually excluding sets of variables and can be treated individually. We derive mutatis mutandis exactly the same computations as for the real case. We obtain
\begin{align*}
    f(H)={}& \exp \bigg( -\sum_{i_1\leq\dots\leq i_p} \frac{a}{\Gamma_{i_1,\dots,i_p} } \bigl(H^{(0)}_{i_{1},\dots,i_p}\bigr)^2 - \sum_{\genfrac{}{}{0pt}{}{i_1\leq \dots \leq i_p }{\text{not paired}} } \frac{a}{\Gamma_{i_1,\dots,i_p}} \bigl(H^{(1)}_{i_{1},\dots,i_p}\bigr)^2 \\
    & + \sum_{i_1 \leq\dots \leq i_{p/2}} \frac{b}{\Gamma_{i_1,i_1,\dots,i_{p/2},i_{p/2}}} H^{(0)}_{i_1,i_1,\dots,i_{p/2},i_{p/2}} +c \bigg) \\
    ={}& \exp \left( -a \sum_{i_1,\dots,i_p} H_{i_{1},\dots,i_p} \bar H_{i_1,\dots,i_p} + b \sum_{i_1,\dots, i_{p/2}} H_{i_1,i_1,\dots,i_{p/2},i_{p/2}} +c \right)
\end{align*}
in the unitary case, and
\begin{align}
    f(\mathrm{H}) ={}& \exp \bigg( -\sum_{\epsilon \in \mathcal{E}} \sum_{i_1\leq\dots\leq i_p} \frac{a}{\Gamma_{i_1,\dots,i_p}} \bigl(Q^{(\epsilon)}_{i_{1},\dots,i_p}\bigr)^2 -  \sum_{\epsilon \notin \mathcal{E}} \sum_{\genfrac{}{}{0pt}{}{i_1\leq \dots \leq i_p }{\text{not paired}} } \frac{a}{\Gamma_{i_1,\dots,i_p}} \bigl(Q^{(\epsilon)}_{i_{1},\dots,i_p}\bigr)^2 \nonumber\\
    &  + \sum_{i_1 \leq\dots \leq i_{p/2}} \frac{b}{\Gamma_{i_1,i_1,\dots,i_{p/2},i_{p/2}}} Q^{(\epsilon)}_{i_1,i_1,\dots,i_{p/2},i_{p/2}} +c \bigg) \label{eq-B}
\end{align}
in the symplectic one. Since
\begin{align*}
    \| Q_{i_1,\dots,i_p} \|^2_{\mathrm{F}} &= \sum_{\iota_1,\dots,\iota_p \in \{1,2\} } \vert(Q_{i_1,\dots,i_p})_{\iota_1,\dots,\iota_p}\vert^2
    \\
    &= \sum_{\iota_1,\dots,\iota_p \in \{1,2\} } \mathrm{Re}((Q_{i_1,\dots,i_p})_{\iota_1,\dots,\iota_p})^2 + \mathrm{Im}((Q_{i_1,\dots,i_p})_{\iota_1,\dots,\iota_p})^2
    \\
    &= \sum_{\epsilon: n_1(\epsilon)+n_3(\epsilon) \equiv 0 [2] } \bigl(Q^{(\epsilon)}_{i_1,\dots,i_p}\bigr)^2 + \sum_{\epsilon: n_1(\epsilon)+n_3(\epsilon) \equiv 1 [2] } \bigl(Q^{(\epsilon)}_{i_1,\dots,i_p}\bigr)^2 \\
    &= \sum_{\epsilon} \bigl(Q^{(\epsilon)}_{i_1,\dots,i_p}\bigr)^2,
\end{align*}
then
\begin{align*}
    \| \mathrm{H} \|^2_{\mathrm{F}} = \sum_{i_1,\dots,i_p } \| Q_{i_1,\dots,i_p} \|^2_{\mathrm{F}} = \sum_{i_1,\dots,i_p } \sum_{\epsilon } \bigl(Q^{(\epsilon)}_{i_1,\dots,i_p}\bigr)^2,
\end{align*}
and we deduce immediately from equation \eqref{eq-B} the desired result.

\subsection{Letac extension}\label{sec:let}

In this subsection, we discuss briefly about a Letac-type extension of the theorem. We say that a~real symmetric tensor $H$ is {\em isotropic} if $H / \| H \|_{\mathrm{F}}$ is uniformly distributed on the unit Frobenius sphere of $\mathcal{S}^{(p)}(N)$. In the case $p=2$, such matrices have been studied by Kopp and Miller in~\cite{koppmiller} as spherical matrix ensembles, but we would prefer the term isotropy instead of sphericity which could refer to rotational invariance in the physics literature. The original result for vectors by Letac is the following.
\begin{Proposition}[Letac \cite{letac}]\label{prop:let}
    A random vector $H \in \mathbb{R}^N$, $N \geq 3$ has independent coordinates and is isotropic if and only if its components are Gaussian, centered, with the same variance~${\gamma \in \mathbb{R}^*_+}$.
\end{Proposition}
\begin{Corollary}\label{cor:let}
    A random vector $H \in \mathcal{S}^{(p)}(N)$ with $p\geq 2, N \geq 2$ has independent coordinates up to symmetries and is isotropic if and only if it belongs to the $\mathrm{GOTE}(0,\gamma)$, $\gamma \in \mathbb{R}^*_+$.
\end{Corollary}
\begin{proof}
Let $p\geq 2,N\geq 2$, $K:=\sum_{i_1=1}^N \sum_{i_2=i_1}^N \dots \sum_{i_p=i_{p-1}}^N 1$ and $\Phi\colon\mathcal{S}^{(p)}(N)\to\mathbb{R}^{K}$ be defined~by
\[
\Phi(H)=\Bigr(\sqrt{\Gamma_{i_1,\dots,i_p}} H_{i_1,\dots,i_p}, 1\leq i_1 \leq \dots \leq i_p \leq N\Bigr).
\]
Then $\Phi$ is linear, bijective, and an isometry in the sense that $|\Phi(H)|=\|H\|_{\mathrm{F}}$ for all ${H\in\mathcal{S}^{(p)}(N)}$. Next, $\Phi(H)$ is a random vector of $\mathbb{R}^K$, $K\geq3$, with independent coordinates since $\Phi$ is a~tensor function. Moreover, since $H/\|H\|_{\mathrm{F}}$ is uniformly distributed on the unit Frobenius sphere of~$\mathcal{S}^{(p)}(N)$, and since $\Phi$ is a linear isometry, it follows that $\Phi(H)/|\Phi(H)|=\Phi(H/\|H\|_{\mathrm{F}})$ is uniformly distributed on the unit Euclidean sphere of $\mathbb{R}^K$. Therefore, by Proposition \ref{prop:let}, $\Phi(H)$ is Gaussian, and therefore $H$ is Gaussian since $\Phi$ is linear and bijective.
\end{proof}

The question is now: Can we deduce Maxwell theorem from this result as it is the case for vectors when $N\geq 3$? Note that for $p$ even and for all $\alpha\in\mathbb{R}$, $\alpha \mathcal{I}$ is orthogonal invariant, but~${\alpha \mathcal{I}/\|\alpha \mathcal{I}\|_{\mathrm{F}}=\mathcal{I}/\sqrt{\vert \mathfrak{P}\vert}}$ is not uniformly distributed on the unit Frobenius sphere. Moreover, if the law of $H$ is orthogonal invariant, then for all $\alpha\in\mathbb{R}$, the law of $H+\alpha I$ is also orthogonal invariant. We conjecture that if $H$ is orthogonal invariant with independent entries, then $H':=S-\mathbb{E}(S_{1,\dots,1}) \mathcal{I}$ is isotropic. This would give the theorem by applying Corollary~\ref{cor:let} to $H'$. Remark that in the matrix case, if $H$ is orthogonal invariant with independent entries, $\mathbb{E}(H_{ij})=0$ for all~${1\leq i<j\leq n}$ and the law of $H_{kk}$ does not depend on $k$ so $H'$ is centered, orthogonal invariant with independent entries. Is it sufficient for isotropy ? We leave this question open to the interested reader.

\section{Complete family of invariants}\label{basis}

We define properly the {\em trace invariants} and then prove that any invariant polynomial can be written as a linear combination of the trace invariants. This name was first given by Gurau in~\cite{gurauarticle14}. Let \smash{$\mathcal{G}_n^{(p)}$} the set of all connected $p$-regular multigraphs with $n$ vertices, that is, graphs where multiedges and self-loops are allowed and each vertex has multiplicity $p$, with a given order for the $p$ edges around each vertex. We denote $\partial v$ the $p$-tuple of neighboring edges of $v$. The edges are indexed by \smash{$i_1,\dots,i_{np/2}$}. Note than $np$ must be even, otherwise there is no trace invariant of rank $n$. Then, for $H$ a tensor of order $p$, dimension $N$ and \smash{$G=(V,E) \in \bigsqcup_{n\geq 0} \mathcal{G}_n^{(p)}$}, we define the trace invariant associated to $G$ as the polynomial
\[ \mathfrak{m}_G(H)= \sum_{1 \leq i_1,\dots,i_E \leq N} \prod_{v=1}^N H_{i_{(\partial v)_1,\dots,(\partial v)_p}}. \]
In the hermitian and self-dual hermitian cases, we ask moreover that the edges of the multigraphs connect two halfedges of different parity, that is, $(\partial v)_{2t-1}$ and $(\partial w)_{2k}$ for two vertices $v$ and $w$ no necessarily distinct.
\begin{Example}\label{example2} Let $H$ be a (symmetric or hermitian) tensor of order $4$ and dimension~$N$. We~give an explicit example of trace invariant associated to a $4$-regular multigraph.
\begin{figure}[ht]
    \centering
\begin{tikzpicture}[scale=0.5]
    \draw[fill=black] (0,0) circle (5pt);
    \filldraw[black] (0,-4) circle (5pt);
    \filldraw[black] (6,-2) circle (5pt);
    \draw (0,0) node[left] {$u$};
    \draw (0,-4) node[left] {$v$};
    \draw (6,-2) node[above] {$w$};
    \draw (0,0) -- (0,-4) ;
    \draw (0,0) .. controls (-1,-1) and (-1,-3) .. (0,-4);
    \draw (0,0) .. controls (1,-1) and (1,-3) .. (0,-4);
    \draw (-1,-2) node {$1$};
    \draw (0,-2) node {$2$};
    \draw (1,-2) node {$3$};
    \draw (3,-1) node {$4$};
    \draw (3,-3) node {$5$};
    \draw (8,-2) node {$6$};
    \draw (0,0) -- (6,-2);
    \draw (0,-4) -- (6,-2);
    \draw (6,-2) .. controls (8,-4) and (8,0) .. (6,-2);
\end{tikzpicture}
\caption{For this graph, we have $\partial u =(1,2,3,4), \partial v =(5,3,2,1)$ and $\partial w =(4,5,6,6)$. We decorate the vertices $u$, $v$ and $w$ by $H$ and each edge by an index ($i_k$ for edge $k$). The associated trace invariant is given by \smash{$\sum_{1 \leq i_1,\dots,i_6 \leq N} H_{i_1,i_2,i_3,i_4} H_{i_5,i_3,i_2,i_1} H_{i_4,i_5,i_6,i_6}$}.}
\end{figure}
\end{Example}

We claim that the trace invariant are a complete family of all the invariant polynomials in both three cases. Note that exactly the same result holds for polynomials in the entries of a~family of tensors, with possibly different orders. A complete family is given by trace invariants associated to graphs having vertices with multiplicity matching with the order of a tensor of the family (these polynomials are also called tensor maps or tensor networks).

\subsection{Orthogonal case}

The proof in the orthogonal case is already properly given by Kunisky, Moore and Wein in \cite[Appendix A]{kunmoorewein}. They recall references where the result appeared \cite{goodmanwallach,wey66} but they give a proof more suitable to our context. We recall here their path. Let $H$ be a real symmetric tensor of order $p$, dimension $N$ and $R$ be a polynomial of degree $n$ in the entries of $H$. Then there exists a vector of coefficients $C \in (\mathbb{R}^{N^p})^{n}\simeq\bigl(\mathbb{R}^N\bigr)^{\otimes np}$ such that
\[R(H)= \sum_{i_1,\dots,i_{np}} \underbrace{H_{i_1,\dots,i_p}\dots H_{i_{(n-1)p+1},\dots,i_{np}}}_{=(H^{\otimes n})_{i_1,\dots,i_{np}}} C_{i_1,\dots,i_{np}} =: \big\langle H^{\otimes n}, C \big\rangle. \]
Then for $U \in \mathrm{O}(N)$, since $R$ is orthogonal invariant, we have
\[ R(H) = R\bigl(U^{\mathsf T} \cdot H\bigr)= \big\langle U^{\mathsf T} \cdot H^{\otimes n}, C \big\rangle = \big\langle H^{\otimes n}, U \cdot C \big\rangle = \big\langle H^{\otimes n}, \bigl(U^{\otimes np}\bigr)^{\sigma} \star_{np} C \big\rangle, \]
with $\sigma = (2 \ \ np+1)(4 \ \ np+3)\cdots (np \ \ 2np-1)$, so $(U^{\otimes np, \sigma})_{i_1,\dots,i_{2np}}=\prod_{t=1}^{np} U_{i_t,i_{np+t}} $ (we contract along the odd legs because each $U$ has one contracted leg in $U \cdot C$). We can now symmetrize $C$ by taking the expectation over the Haar measure, that is,
\[ R(H) = \big\langle H^{\otimes n}, \mathbb{E}_{U \in \mathrm{O}(N)} U^{\otimes np, \sigma} \star_{np} C \big\rangle. \]
It is well known \cite{collinssniady04,weingarten} that the expectation over the orthogonal group can be written in terms of the Weingarten functions, that is,
\[ \bigl(\mathbb{E}_{U \in \mathrm{O}(N)} U^{\otimes np, \sigma}\bigr)_{i_1,\dots,i_{2np}} = \sum_{\nu,\tau \in \mathcal{M}(np)} \mathrm{Wg}_{\nu,\tau} \prod_{t=1}^{np} \delta_{i_t,i_{\nu(t)}} \delta_{i_{np+t},i_{np+\tau(t)}}, \]
where $\mathcal{M}(np) \in S_{np}$ is the set of involutions without fixed point (perfect matchings) and ${\mathrm{Wg}_{\nu,\tau} \in \mathbb{R}}$ is the Weingarten coefficient associated to $(\nu,\tau)$. Now we get
\[
\bigl(\mathbb{E}_{U \in \mathrm{O}(N)} U^{\otimes np, \sigma} \star_{np} C\bigr)_{i_1,\dots,i_{np}} = \sum_{\tau \in \mathcal{M}(np)} \alpha(\tau) \prod_{t=1}^{np} \delta_{i_t,i_{\tau(t)}},
\]
and denoting $S_{\tau}$ the tensor associated to the matching $\tau$ with entries $(S_{\tau})_{i_1,\dots,i_p}=\prod_{t=1}^{np} \delta_{i_t,i_{\tau(t)}}$, we have
\[
R(H) = \sum_{\tau \in \mathcal{M}(np)} \alpha(\tau) \big\langle H^{\otimes n}, S_\tau \big\rangle = \sum_{\tau \in \mathcal{M}(np)} \alpha(\tau) \mathfrak{m}_{G(\tau)}(H),
 \]
where $G(\tau)$ is the multigraph canonically associated to $\tau$, that is, a vertex $v \in \{ 0,\dots,n-1 \}$ has neighboring edges $\partial v =(vp+1,\dots,(v+1)p)$ and the halfedge $(\partial v)_k$ is matched with $\tau((\partial v)_k)$. This concludes the proof.

\subsection{Unitary case}

Let us adapt this in the unitary case. Let $p$ be an even integer and $H$ a $p$-order hermitian tensor. The computation is slightly different from the one did for the case of {\em local} unitary invariance, where the legs are distinguished in $p$ different colors and cannot be matched with one of another color, see for instance the rich monograph of Gurau \cite{guraubook}. Let $R$ be an invariant polynomial of degree $n$. We still have $R(H) = \big\langle H^{\otimes p}, C \big\rangle$, so for $U \in \mathrm{U}(N)$,
\begin{align*}
R(H) & = R\bigl(U^{\mathsf T} \cdot H\bigr)= \big\langle U^{\mathsf T} \cdot H^{\otimes n}, C \big\rangle\\
& = \big\langle H^{\otimes n}, U \cdot C \big\rangle = \big\langle H^{\otimes n}, \bigl(U^{\otimes np/2} \otimes \bar U^{\otimes np/2}\bigr)^{\sigma} \star_{np} C \big\rangle,
\end{align*}
where we denote
\begin{itemize}\itemsep=0pt
    \item $\sigma = (2 \ \ np+1)(4 \ \ np+3)\cdots (np \ \ 2np-1)$, and
    \item \smash{$\bigl(U^{\otimes np/2} \otimes \bar U^{\otimes np/2}\bigr)^{\sigma}_{i_1,\dots,i_{2np}}=\prod_{t=1}^{np/2} U_{i_{2t-1},i_{np+2t-1}} \bar U_{i_{2t},i_{np+2t}} $}.
\end{itemize}
Taking the expectation over the Haar measure gives
\[ R(H) = \big\langle H^{\otimes n}, \mathbb{E}_{U \in \mathrm{U}(N)} \bigl(U^{\otimes np/2} \otimes \bar U^{\otimes np/2}\bigr)^{\sigma} \star_{np} C \big\rangle, \]
and we derive thanks to complex Weingarten calculus developed by Collins et al.\ \cite{collinssniady04}
\begin{gather*}
 \bigl(\mathbb{E}_{U \in \mathrm{U}(N)} \bigl(U^{\otimes np/2} \otimes \bar U^{\otimes np/2}\bigr)^{\sigma}\bigr)_{i_1,\dots,i_{2np}}\\
  \qquad= \sum_{\nu, \tau \in \mathcal{M}(\{1,3,\dots,np-1\}, \{2,4,\dots,np\})} \mathrm{Wg}_{\sigma,\tau} \prod_{t=1}^{np} \delta_{i_t,i_{\nu(t)}} \delta_{i_{np+t},i_{np+\tau(t)}},
  \end{gather*}
where $\mathcal{M}(\{1,3,\dots,np-1\}, \{2,4,\dots,np\})$ is the set of involution with no fixed points from $\{1,\dots,np\}$ into itself such that all the images of elements of $\{1,3,\dots,np-1\}$ are in $\{2,4,\dots,np\}$ and reciprocally ($\mathcal{M}$ is for ``matching''). Hence,
\[R(H) = \sum_{\tau \in \mathcal{M}(\{1,3,\dots,np-1\}, \{2,4,\dots,np\})} \alpha(\tau) \mathfrak{m}_{G(\tau)}(H), \]
where $G(\tau)$ is the multigraph associated to $\tau$, where the even halfedges are then matched with odd halfedges.
\begin{figure}[ht]
    \centering
\begin{tikzpicture}[scale=0.45]
    \draw (-4,0) node {$R(H)=$};
    \draw[fill=black] (0,4) circle (5pt);
    \draw[fill=black] (0,-4) circle (5pt);
    \draw[black,fill=gray] (0,0) circle (8pt);
    \draw (0.3,0) node[right] {$C$};
    \draw (0,4) node[right] {$H$};
    \draw (0,-4) node[right] {$H$};
    \draw (0,0) .. controls (-2,1) and (-2,3) .. (0,4);
    \draw (0,0) .. controls (2,1) and (2,3) .. (0,4);
    \draw (0,0) .. controls (-1,1) and (-1,3) .. (0,4);
    \draw (0,0) .. controls (1,1) and (1,3) .. (0,4);
    \draw (0,0) .. controls (-2,-1) and (-2,-3) .. (0,-4);
    \draw (0,0) .. controls (2,-1) and (2,-3) .. (0,-4);
    \draw (0,0) .. controls (-1,-1) and (-1,-3) .. (0,-4);
    \draw (0,0) .. controls (1,-1) and (1,-3) .. (0,-4);
    \draw (4,0) node {$=$};
    \draw[fill=black] (8,4) circle (5pt);
    \draw[fill=black] (8,-4) circle (5pt);
    \draw[black,fill=gray] (8,0) circle (8pt);
    \draw (8.3,0) node[right] {$C$};
    \draw (8,4) node[right] {$H$};
    \draw (8,-4) node[right] {$H$};
    \draw (8,0) -- (6,2);
    \draw (8,4) -- (6,2);
    \draw (8,0) -- (10,2);
    \draw (8,4) -- (10,2);
    \draw (8,0) -- (7,2);
    \draw (8,4) -- (7,2);
    \draw (8,0) -- (9,2);
    \draw (8,4) -- (9,2);
    \draw (8,0) -- (6,-2);
    \draw (8,-4) -- (6,-2);
    \draw (8,0) -- (10,-2);
    \draw (8,-4) -- (10,-2);
    \draw (8,0) -- (7,-2);
    \draw (8,-4) -- (7,-2);
    \draw (8,0) -- (9,-2);
    \draw (8,-4) -- (9,-2);
    \draw[fill=black] (6,2) circle (3pt);
    \draw[fill=black] (10,2) circle (3pt);
    \draw[fill=black] (7,2) circle (3pt);
    \draw[fill=black] (9,2) circle (3pt);
    \draw[fill=black] (6,-2) circle (3pt);
    \draw[fill=black] (10,-2) circle (3pt);
    \draw[fill=black] (7,-2) circle (3pt);
    \draw[fill=black] (9,-2) circle (3pt);
    \draw (6,2) node[left] {$U$};
    \draw (10,2) node[right] {$\bar{U}$};
    \draw (7,2) node[right] {$\bar{U}$};
    \draw (9,2) node[left] {$U$};
    \draw (6,-2) node[left] {$U$};
    \draw (10,-2) node[right] {$\bar{U}$};
    \draw (7,-2) node[right] {$\bar{U}$};
    \draw (9,-2) node[left] {$U$};
    \draw (12,0) node {$=$};
    \draw[fill=black] (16,4) circle (5pt);
    \draw[fill=black] (16,-4) circle (5pt);
    \draw[black,fill=gray] (16,0) circle (8pt);
    \draw (16.3,0) node[right] {$\bigl(U^{\otimes np/2} \otimes \bar U^{\otimes np/2}\bigr)^{\sigma} \star_{np} C$};
    \draw (16,4) node[right] {$H$};
    \draw (16,-4) node[right] {$H$};
    \draw (16,0) .. controls (14,1) and (14,3) .. (16,4);
    \draw (16,0) .. controls (18,1) and (18,3) .. (16,4);
    \draw (16,0) .. controls (15,1) and (15,3) .. (16,4);
    \draw (16,0) .. controls (17,1) and (17,3) .. (16,4);
    \draw (16,0) .. controls (14,-1) and (14,-3) .. (16,-4);
    \draw (16,0) .. controls (18,-1) and (18,-3) .. (16,-4);
    \draw (16,0) .. controls (15,-1) and (15,-3) .. (16,-4);
    \draw (16,0) .. controls (17,-1) and (17,-3) .. (16,-4);
    \draw (-3,-10) node {$=$};
    \draw[fill=black] (0,-6) circle (5pt);
    \draw[fill=black] (0,-14) circle (5pt);
    \draw[black,fill=gray] (0,-10) circle (8pt);
    \draw (0.3,-10) node[right] {$\mathbb{E} \bigl(U^{\otimes np/2} \otimes \bar U^{\otimes np/2}\bigr)^{\sigma} \star_{np} C$};
    \draw (0,-6) node[right] {$H$};
    \draw (0,-14) node[right] {$H$};
    \draw (0,-10) .. controls (-2,-9) and (-2,-7) .. (0,-6);
    \draw (0,-10) .. controls (2,-9) and (2,-7) .. (0,-6);
    \draw (0,-10) .. controls (-1,-9) and (-1,-7) .. (0,-6);
    \draw (0,-10) .. controls (1,-9) and (1,-7) .. (0,-6);
    \draw (0,-10) .. controls (-2,-11) and (-2,-13) .. (0,-14);
    \draw (0,-10) .. controls (2,-11) and (2,-13) .. (0,-14);
    \draw (0,-10) .. controls (-1,-11) and (-1,-13) .. (0,-14);
    \draw (0,-10) .. controls (1,-11) and (1,-13) .. (0,-14);
    \draw (12,-10) node {$=$};
    \draw (12,-12) node {$\downarrow$};
    \draw (12,-14) node {Weingarten calculus};
    \draw (14,-10) node {$\alpha_1$};
    \draw[fill=black] (16,-8) circle (5pt);
    \draw[fill=black] (16,-12) circle (5pt);
    \draw (16,-8) node[right] {$H$};
    \draw (16,-12) node[right] {$H$};
    \draw (16,-8) .. controls (14,-9) and (16,-11) .. (16,-12);
    \draw (16,-8) .. controls (15,-9) and (17,-11) .. (16,-12);
    \draw (16,-8) .. controls (16,-9) and (18,-11) .. (16,-12);
    \draw (16,-8) .. controls (19,-9) and (13,-11) .. (16,-12);
    \draw (18,-10) node {$+$};
    \draw (20,-10) node {$\alpha_2$};
    \draw[fill=black] (22,-8) circle (5pt);
    \draw[fill=black] (22,-12) circle (5pt);
    \draw (22,-8) node[right] {$H$};
    \draw (22,-12) node[right] {$H$};
    \draw (22,-8) .. controls (20,-10) and (22,-10) .. (22,-8);
    \draw (22,-12) .. controls (22,-10) and (24,-10) .. (22,-12);
    \draw (22,-8) .. controls (23,-9) and (21,-11) .. (22,-12);
    \draw (22,-8) .. controls (25,-9) and (19,-11) .. (22,-12);
    \draw (25,-10) node {$+\cdots$};
\end{tikzpicture}
    \caption{Sketch of proof, unitary case.}    \label{fig:traceunitary}
\end{figure}

\subsection{Symplectic case}

Let us now treat the symplectic case. Recall that $J$ is the canonical symplectic form corresponding to the matrix \[J=\mathbf{e_2}I^{\mathbb{H}}_N =\begin{pmatrix} 0 & -1 & 0 & \dots &0 \\ 1 & 0 & -1 & \dots & 0 \\ 0 & \ddots & \ddots & \ddots & \vdots \\ \vdots & \ddots & \ddots & \ddots & -1 \\ 0 & \dots & \dots & 1 & 0 \end{pmatrix} =-J^{\mathsf T}.\]
With the considered similarity transformation, we have for $R$ an invariant polynomial of degree~$n$ and $U \in \mathrm{Sp}(2N)$,
\[ R(H) = R\bigl(U^{\mathsf T} \cdot H\bigr)= \big\langle H^{\otimes n}, \bigl(U^{\otimes np/2} \otimes (-JUJ)^{\otimes np/2}\bigr)^{\sigma} \star_{np} C \big\rangle, \]

\noindent
where we denote{\samepage
\begin{itemize}\itemsep=0pt
    \item $\sigma = (2 \ \ np+1)(4 \ \ np+3)\cdots (np \ \ 2np-1)$, and
    \item \smash{$\bigl(U^{\otimes np/2} \otimes (-JUJ)^{\otimes np/2}\bigr)^{\sigma}_{i_1,\dots,i_{2np}} = \prod_{t=1}^{np/2} U_{i_{2t-1},i_{np+2t-1}} (-JUJ)_{i_{2t},i_{np+2t}} $}.
\end{itemize}}%
Again taking the expectation over the Haar measure gives
\[ R(H) = \big\langle H^{\otimes n}, \mathbb{E}_{U \in \mathrm{Sp}(2N)} \bigl(U^{\otimes np/2} \otimes (-JUJ)^{\otimes np/2}\bigr)^{\sigma} \star_{np} C \big\rangle, \]
and we derive thanks to symplectic Weingarten calculus \cite{collinssniady04}
\begin{gather*}
    \bigl( \mathbb{E}_{U \in \mathrm{Sp}(2N)} \bigl(U^{\otimes np/2} \otimes (-JUJ)^{\otimes np/2}\bigr)^{\sigma}\bigr)_{i_1,\dots,i_{2np}} \\
    \qquad= (-1)^{np/2} \sum_{k_1,k'_2,\dots,k_{np-1},k'_{np}} \mathbb{E}_{U \in \mathrm{Sp}(2N)}\prod_{t=1}^{np/2} U_{i_{2t-1},i_{np+2t-1}} J_{i_{2t},k_{2t-1}} U_{k_{2t-1},k'_{2t}} J_{k'_{2t},i_{np+2t}}\\
    \qquad= \sum_{k_i,k'_i} \sum_{\nu,\tau \in \mathcal{M}(\{1,3,\dots,np-1\}, \{2,4,\dots,np\})} \mathrm{Wg}_{\nu,\tau}\\
     \phantom{\qquad= }{}\times\prod_{t=1}^{np/2} J_{i_{2t-1},k_{\nu(2t)}} J_{i_{np+2t-1},k'_{\tau(2t-1)}} J_{i_{2t},k_{2t-1}} J_{k'_{2t},i_{np+2t}} \\
    \qquad= \sum_{\nu,\tau \in \mathcal{M}(\{1,3,\dots,np-1\}, \{2,4,\dots,np\})} \mathrm{Wg}_{\nu,\tau}\\
     \phantom{\qquad= }{}\times\prod_{t=1}^{np/2} \sum_{k} J_{i_{2t-1},k}  J_{i_{\nu(2t-1)},k} \sum_{k'} J_{i_{np+2t-1},k'} J_{k',i_{np+\tau(2t-1)}} \\
    \qquad= \sum_{\nu,\tau \in \mathcal{M}(\{1,3,\dots,np-1\}, \{2,4,\dots,np\})} \mathrm{Wg}_{\nu,\tau} \prod_{t=1}^{np/2} \delta_{i_{2t-1},i_{\nu(2t-1)}} \delta_{i_{np+2t-1,i_{np+\tau(2t-1)}}}. \\
    \qquad= \sum_{\nu,\tau \in \mathcal{M}(\{1,3,\dots,np-1\}, \{2,4,\dots,np\})} \mathrm{Wg}_{\nu,\tau} \prod_{t=1}^{np} \delta_{i_t,i_{\nu(t)}} \delta_{i_{np+t,i_{np+\tau(t)}}}.
\end{gather*}
So finally, as in the hermitian case, we find
\[R(H) = \sum_{\tau \in \mathcal{M}(\{1,3,\dots,np-1\}, \{2,4,\dots,np\})} \alpha(\tau) \mathfrak{m}_{G(\tau)}(H), \]
where $G(\tau)$ is the multigraph associated to $\tau$, where the even halfedges are then matched with odd halfedges.

\subsection{The melon graph}\label{melonfrob}

The melon is the graph with two vertices linked by $p$ edges, in the hermitian case the $t$-th halfedge of the first vertex is linked with the $(t+1)$-th one of the second vertex and in the self-dual hermitian case the $2t$-th halfedge of a vertex is linked with the $(2t-1)$-th one of the other vertex. Then, the computation of the trace invariant is straightforward:
\begin{itemize}\itemsep=0pt
    \item[--] in the real symmetric case,
    \smash{$\sum_{i_1,\dots,i_p} H_{i_1,\dots,i_p} H_{i_1,\dots,i_p} = \| H \|_{\mathrm{F}}^2$},
    \item[--] in the hermitian case,
    \begin{align*}
    \sum_{i_1,\dots,i_p} H_{i_1,\dots,i_p} H_{i_2,\dots,i_p,i_1} ={}& \sum_{i_1,\dots,i_p} \bigl(H^{(0)}_{i_1,\dots,i_p} +i H^{(1)}_{i_1,\dots,i_p} \bigr) \\
    &\times\bigl(H^{(0)}_{i_1,\dots,i_p} + i \underbrace{\epsilon((1 \ \ 2 \dots p))}_{=(-1)^{p-1}=-1} H^{(1)}_{i_1,\dots,i_p}\bigr)
    = \| H \|_{\mathrm{F}}^2,
    \end{align*}
    \item[--] in the self-dual hermitian case, let $\iota_k=i_k-2\lfloor\frac{i_k}{2}\rfloor+1 \in \{1,2\}$,
\begin{align*}
    \sum_{1\leq i_1,\dots,i_{p}\leq 2N} &  H_{i_1,\dots,i_p} H_{i_2,i_1,\dots,i_p,i_{p-1}} \\
    &= \sum_{i_1,\dots,i_{p} }  \bigl(Q_{\lfloor \frac{i_{1}}{2} \rfloor,\dots,\lfloor \frac{i_{p}}{2} \rfloor}\bigr)_{\iota_1,\dots,\iota_p} \bigl(Q_{\lfloor \frac{i_{2}}{2} \rfloor,\lfloor \frac{i_{1}}{2} \rfloor,\dots,\lfloor \frac{i_{p}}{2} \rfloor,\lfloor \frac{i_{p-1}}{2} \rfloor}\bigr)_{\iota_2,\iota_1,\dots,\iota_p,\iota_{p-1}} \\
    &= \sum_{i_1,\dots,i_{p} } \sum_{\epsilon \in \mathcal{E}} \bigl(Q^{\epsilon}_{\lfloor \frac{i_{1}}{2} \rfloor,\dots,\lfloor \frac{i_{p}}{2} \rfloor}\bigr)^2 \biggl(\bigotimes_{s=1}^{p/2} \mathbf{e_{\epsilon_s}} \biggr)_{\iota_1,\dots,\iota_p} \biggl(\bigotimes_{s=1}^{p/2} \mathbf{e_{\epsilon_s}}\biggr)_{\iota_2,\iota_1,\dots,\iota_p,\iota_{p-1}} \\ & \hspace{1cm} + \sum_{\epsilon \notin \mathcal{E}} (-1)^{p/2} \bigl(Q^{\epsilon}_{\lfloor \frac{i_{1}}{2} \rfloor,\dots,\lfloor \frac{i_{p}}{2} \rfloor}\bigr)^2 \biggl(\bigotimes_{s=1}^{p/2} \mathbf{e_{\epsilon_s}}\biggr)_{\iota_1,\dots,\iota_p} \biggl(\bigotimes_{s=1}^{p/2} \mathbf{e_{\epsilon_s}}\biggr)_{\iota_2,\iota_1,\dots,\iota_p,\iota_{p-1}} \\
    &= \sum_{i_1,\dots,i_{p} } \sum_{\epsilon \in \mathcal{E}} (-1)^{n_2}  \bigl(Q^{\epsilon}_{\lfloor \frac{i_{1}}{2} \rfloor,\dots,\lfloor \frac{i_{p}}{2} \rfloor}\bigr)^2 -  \sum_{\epsilon \notin \mathcal{E}} (-1)^{n_2} \bigl(Q^{\epsilon}_{\lfloor \frac{i_{1}}{2} \rfloor,\dots,\lfloor \frac{i_{p}}{2} \rfloor}\bigr)^2,
\end{align*}
where $(-1)^{p/2}=-1$ because $p \equiv 2 \pmod 4$. Remark that by choice of~$\mathcal{E}$ as the set of~$\epsilon$ such that $n_1(\epsilon)+n_3(\epsilon)\equiv n_2(\epsilon) \pmod 2$, where $n_1+n_3$ is the number of imaginary quaternions and $n_2$ the number of antisymmetric ones appearing in the tensor product of~$\mathbf{e_{\epsilon_s}}$, we have
\begin{align*}
    \bar H_{i_1,\dots,i_p} &= \sum_{\epsilon\colon n_1 \equiv n_3 [2]} Q^{\epsilon}_{\lfloor \frac{i_{1}}{2} \rfloor,\dots,\lfloor \frac{i_{p}}{2} \rfloor} - \sum_{\epsilon\colon n_1 \not\equiv n_3[2]}  Q^{\epsilon}_{\lfloor \frac{i_{1}}{2} \rfloor,\dots,\lfloor \frac{i_{p}}{2} \rfloor} \\
    & = \sum_{\epsilon\colon  n_1 \equiv n_3[2],\, n_2\equiv 0[2]} Q_{\lfloor \frac{i_{1}}{2} \rfloor,\dots,\lfloor \frac{i_{p}}{2} \rfloor}^{\epsilon} + \sum_{\epsilon\colon n_1 \equiv n_3[2],\, n_2 \equiv 1[2]} Q_{\lfloor \frac{i_{1}}{2} \rfloor,\dots,\lfloor \frac{i_{p}}{2} \rfloor}^{\epsilon} \\
    & \hspace{1cm} - \sum_{\epsilon\colon n_1 \not\equiv n_3[2],\, n_2\equiv 0[2]} Q_{\lfloor \frac{i_{1}}{2} \rfloor,\dots,\lfloor \frac{i_{p}}{2} \rfloor}^{\epsilon} - \sum_{\epsilon\colon n_1 \not\equiv n_3 [2],\, n_2 \equiv 1[2]} Q_{\lfloor \frac{i_{1}}{2} \rfloor,\dots,\lfloor \frac{i_{p}}{2} \rfloor}^{\epsilon} \\
    & =\sum_{\epsilon \in \mathcal{E}} (-1)^{n_2} Q_{\lfloor \frac{i_{1}}{2} \rfloor,\dots,\lfloor \frac{i_{p}}{2} \rfloor}^{\epsilon} - \sum_{\epsilon \notin \mathcal{E}} (-1)^{n_2} Q_{\lfloor \frac{i_{1}}{2} \rfloor,\dots,\lfloor \frac{i_{p}}{2} \rfloor}^{\epsilon},
\end{align*}
so finally we find
\[ \sum_{1\leq i_1,\dots,i_{p}\leq 2N}  H_{i_1,\dots,i_p} H_{i_2,i_1,\dots,i_p,i_{p-1}}= \sum_{i_1,\dots,i_{p}} H_{i_1,\dots,i_p} \bar H_{i_1,\dots,i_p} = \| H \|_{\mathrm{F}}^2. \]
\end{itemize}

\subsection*{Acknowledgements}

The author would like to thank Djalil Chafa\"{\i} for raising this question and pointing out numerous references during very insightful discussions.

\pdfbookmark[1]{References}{ref}
\LastPageEnding

\end{document}